\documentclass[aps,prb,twocolumn,groupedaddress,showpacs,amsmath]{revtex4}

\usepackage{graphicx}
\usepackage{bm}

\renewcommand{\vec}[1]{\bm{#1}}

\begin{document}

\title{Magnon decay in gapped quantum spin systems}

\author{Alexei Kolezhuk}
\thanks{On leave from the Institute of Magnetism, Academy of Sciences and
  Ministry of Education, 03142 Kiev, Ukraine}
\affiliation{Department of Physics, Harvard University, Cambridge MA
02138}

\author{Subir Sachdev}
\affiliation{Department of Physics, Harvard University, Cambridge MA
02138}

\begin{abstract}
In the O(3) $\sigma$-model description of gapped spin systems,
$S=1$ magnons can only decay into {\em three\/} lower energy
magnons. We argue that the symmetry of the quantum spin
Hamiltonian often allows decay into {\em two\/} magnons, and
compute this decay rate in model systems. Two magnon decay is
present in Haldane gap $S=1$ spin chains, even though it
cannot be induced by any allowed term written in powers and
gradients of the $\sigma$-model field. We compare our results with
recent measurements of Stone {\em et al.} on a two-dimensional
spin system.
\end{abstract}

\maketitle

\emph{Introduction.--} Stone \emph{et al.} \cite{Stone+05} have recently studied the
spectrum of the $S=1$ magnon excitations of the gapped
quasi-two-dimensional spin dimer compound $\rm
(C_{4}H_{12}N_{2})Cu_{2}Cl_{6}$ (piperazinium hexachlorodicuprate,
PHCC). They observed that the magnons become unstable when their
energy crosses the lower boundary of the two-magnon continuum, and
measured magnon lifetimes.
These observations are interesting because, at first glance, they
appear to be in conflict with the standard phenomenological
description of low energy $S=1$ excitations in a confining
spin-gap systems. This model \cite{haldane,affleck} is expressed
in terms of a field ${\bf n}(x, \tau)$, where $x$ is a
$d$-dimensional spatial coordinate, $\tau$ is imaginary time, and
${\bf n}$ is three-component vector in spin space. Using
Landau-Ginzburg arguments, the effective action for the magnons,
near the minimum of their dispersion, can be written in the form
\begin{displaymath}
\mathcal{S}_{\vec{n}} = \int d^d x d \tau \Big\{ \frac{1}{2g} \left(
(\partial_\tau {\vec{n}})^2 + c^2 (\partial_x {\vec{n}})^2 \right) +
\frac{u}{2} \left( {\vec{n}}^2 - \alpha \right)^2 \Big\}
\end{displaymath}
Harmonic quantum fluctuations of ${\vec{n}}$ about ${\vec{n}} = 0$
in a renormalized effective potential then constitute the triply
degenerate magnon in the spin-gap state. In this picture, it is
clear from the quartic non-linearity in $\mathcal{S}_{\vec{n}}$
that this magnon can only decay into three magnons, and decay into
two magnons is prohibited.

It is useful to recall the constraints imposed by symmetry more
carefully for the familiar case of a spin $S$ antiferromagnetic
chain ($d=1$). This model has spin rotation symmetry, time
reversal symmetry (under which $\tau \rightarrow -\tau$ and
$\vec{n} \rightarrow - \vec{n}$), reflection symmetry about sites
(under which $x \rightarrow -x$ and $\vec{n} \rightarrow
\vec{n}$), and reflection symmetry about centers of bonds (under
which, in the continuum limit, $x \rightarrow -x$ and $\vec{n}
\rightarrow -\vec{n}$). These symmetries are very restrictive and
only allow a single term with three powers of $\vec{n}$: the
topological `$\theta$-term'\cite{haldane}
\begin{equation}
\mathcal{S}_\theta = i \theta \int dx d\tau \; {\vec{n}} \cdot
\left(\partial_x {\vec{n}} \times \partial_\tau {\vec{n}} \right),
\label{stheta}
\end{equation}
in the fixed length limit $u \rightarrow \infty$ with $\alpha=1$
and $\theta$ quantized at $\theta=S/2$. In this limit the
integrand in $\mathcal{S}_\theta$ is quantized in integer
multiples of $4 \pi$, and only then is $e^{-
\mathcal{S}_\theta}=1$ invariant under all symmetries noted above.
However, we are interested in the nature of the non-linear terms
as they act on the {\em fully renormalized\/} quasiparticles of
the gapped antiferromagnet with integer $S$. As noted above, these
can be considered to be harmonic fluctuations of a renormalized
field ${\vec{n}}$ about ${\vec{n}} = 0$, and the amplitude of this
renormalized field is {\em not\/} constrained to be unity. For
such a field $\vec{n}$, it is easy to see that there is no term
with three powers of $\vec{n}$, and arbitrary spatial and temporal
gradients which is invariant under the symmetries.

Here we will examine the issue of two magnon decay using a lattice
formulation. Our analysis begins below with a microscopic
Hamiltonian appropriate for PHCC. We will compute the lifetime due
to decay into two magnons, and compare our results with the
observations of Stone {\em et al.}\cite{Stone+05} The
confrontation between theory and experiments places new
constraints on the values of the microscopic exchange constants.
We then show that similar processes also exist for a generic $S=1$
spin chain, and are consistent with all symmetries of the lattice
Hamiltonian; this is the case even though such processes do not
appear in any perturbation of the continuum theory.

\begin{figure}[tb]
\includegraphics[width=45mm]{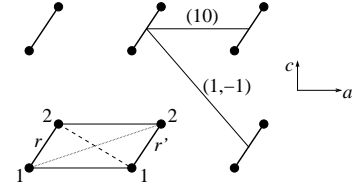}
\caption{\label{fig:dimers}
A schematic view of the coupled spin dimer structure in the
crystallographic $(ac)$ plane of PHCC, and the
notation used. Two spins in each dimer are distinguished by the spin
index $\sigma=1,2$, and each dimer is labeled by the position of its
center $\vec{r}$. Exchange couplings between individual spins are then
denoted as $J^{(\sigma\sigma')}_{\vec{r}-\vec{r}'}$. In PHCC, the most
important exchange links are given by $\vec{r}-\vec{r}'=(1,0)$,
$(0,1)$, $(1,1)$, $(1,-1)$, and possibly also $(2,0)$ and $(0,2)$ (in
lattice units).
}
\end{figure}

\emph{Magnon lifetimes in PHCC.--}
The $\rm Cu^{2+}$ spins in PHCC form a lattice of dimers in the $(ac)$
crystalline plane as shown in Fig.\ \ref{fig:dimers}.
We will denote the interdimer exchange interactions as
$J^{\sigma\sigma'}_{\vec{\delta}}$, where $\sigma,\sigma'=1,2$ label two
spins inside a dimer and  the vector $\vec{\delta}=(m,n)\equiv m \widehat{\vec{a}} +n
\widehat{\vec{c}}$
connects
the dimer centers.
For a
description of such systems it is
convenient to use the so-called
\emph{bond-boson} formalism \cite{SachdevBhatt90} where spin-$\frac{1}{2}$
operators $\vec{S}_{1}$, $\vec{S}_{2}$ at each dimer (see Fig.\
\ref{fig:dimers}) are represented
in terms of three \emph{hardcore} bosons $t_{\alpha}$, $\alpha=(x,y,z)$ which
correspond to three excited triplet states of the dimer:
\begin{equation}
\label{bond-boson}
S_{1,2}^{\alpha}=\pm\frac{1}{2}(t^{\vphantom{\dag}}_{\alpha}+t_{\alpha}^{\dag})
+\frac{1}{2}i\varepsilon_{\alpha\beta\gamma}t^{\vphantom{\dag}}_{\beta}t^{\dag}_{\gamma}.
\end{equation}
If one assumes the isotropic Heisenberg coupling between dimers,
the spin Hamiltonian takes the form
$\widehat{H}=\widehat{H}_{0}+^{H}_{2}+\widehat{H}_{3}+\widehat{H}_{4}$, where
$\widehat{H}_{0}=J_{0}\sum_{\vec{r}}t^{\dag}_{\vec{r}\alpha}t^{\vphantom{\dag}}_{\vec{r}\alpha}$
is determined by the intradimer exchange $J_{0}$, and
$\widehat{H}_{2,3,4}$ correspond to the interdimer
interaction. (The index $\vec{r}$ here labels the sites of the lattice
formed by dimers).

The quadratic part of the interdimer Hamiltonian
\begin{equation}
\label{h2}
\widehat{H}_{2}=\frac{1}{2}\sum_{(\vec{r}\vec{r}')} J^{\rm
  eff}_{\vec{r}-\vec{r}'}(t^{\dag}_{\vec{r}\alpha}t^{\vphantom{\dag}}_{\vec{r}'\alpha} +
  t_{\vec{r}\alpha}t_{\vec{r}'\alpha}+ {\rm h.c.}),
\end{equation}
as well as the four-particle interaction $\widehat{H}_{4}$,
\[
\widehat{H}_{4}=\frac{1}{2}\sum_{(\vec{r}\vec{r}')} J^{\rm
  eff}_{\vec{r}\vec{r}'}\big\{
t^{\dag}_{\vec{r}\alpha}
t^{\dag}_{\vec{r}'\alpha}
t^{\vphantom{\dag}}_{\vec{r}\beta}
 t^{\vphantom{\dag}}_{\vec{r}'\beta}
- t^{\dag}_{\vec{r}'\alpha}
t^{\dag}_{\vec{r}\beta}
t^{\vphantom{\dag}}_{\vec{r}\alpha}
 t^{\vphantom{\dag}}_{\vec{r}'\beta}
\big\},
\]
depend only on specific combinations $J^{\rm eff}_{\vec{r}-\vec{r}'}$
of exchange couplings between
the dimers located at sites $\vec{r}$ and $\vec{r}'$,
\begin{equation}
\label{Jeff}
J^{\rm  eff}_{\vec{r}-\vec{r}'} =
\frac{1}{2}(J_{\vec{r}-\vec{r}'}^{11}
+J_{\vec{r}-\vec{r}'}^{22}
-J_{\vec{r}-\vec{r}'}^{12}
-J_{\vec{r}-\vec{r}'}^{21}).
\end{equation}

The three-magnon term $\widehat{H}_{3}$ depends on
different combinations of exchange constants $F_{\vec{r}-\vec{r}'}$ that are nonzero only if
the interdimer coupling is not symmetric with respect to the
interchange of the spin indices $1$ and $2$:
\begin{flushleft}
\begin{eqnarray}
\label{h3}
&& \widehat{H}_{3}=\frac{i}{2}\sum_{(\vec{r}\vec{r}')}
\varepsilon_{\alpha\beta\gamma}
F_{\vec{r}-\vec{r}'}^{\vphantom{\dag}}
(
t^{\dag}_{\vec{r}\alpha}
t^{\dag}_{\vec{r}'\beta}
t^{\vphantom{\dag}}_{\vec{r}'\gamma} -
t^{\dag}_{\vec{r}'\alpha}
t^{\dag}_{\vec{r}\beta}
t^{\vphantom{\dag}}_{\vec{r}\gamma}
) \! + \!{\rm h.c.} \nonumber\\
&& F_{\vec{r}-\vec{r}'}
= \frac{1}{2}(J_{\vec{r}-\vec{r}'}^{22}-J_{\vec{r}-\vec{r}'}^{11}
-J_{\vec{r}-\vec{r}'}^{12}+J_{\vec{r}-\vec{r}'}^{21}).
\end{eqnarray}
\end{flushleft}
This type of interaction arises generically in the theory; however, it
is usually neglected in practical calculations.

If one neglects magnon interaction completely, taking into account only the
quadratic part of the Hamiltonian
$\widehat{H}_{0}+\widehat{H}_{2}$,
and uses this approximation to describe the
experimentally observed magnon dispersion and to find out the
microscopic exchange
couplings,
one can obtain estimates for the intradimer exchange constant
$J_{0}$ and for ``effective couplings'' $J^{\rm eff}_{\vec{r}-\vec{r}'}$ only. The
individual couplings $J_{\vec{r}-\vec{r}'}^{11}$, $J_{\vec{r}-\vec{r}'}^{12}$, etc. cannot be
determined in this noninteracting magnon approximation that is
commonly used by experimentalists for the description of the
dispersion data (see, e.g., Refs.\
\onlinecite{Leuenberger+84,Kato+98,Stone+01}).

Since an  analytical treatment of the full
interacting Hamiltonian for realistic models does not seem feasible,
one has to resort to extensive numerical calculations to extract the
coupling information from the dispersion data. Such calculations based
on cluster expansions were actually performed for various spin dimer
materials; \cite{MuellerMikeska00,MikeskaMueller02,Oosawa+02} however,
even in this approach the problem remains difficult because for weak
interdimer interactions the magnon
dispersion is rather insensitive to the variations of the individual
exchange couplings which keep the ``effective'' interactions $J^{\rm
  eff}_{\vec{r}-\vec{r}'}$ constant.

The magnon dispersion in PHCC was studied experimentally in Ref.\ \onlinecite{Stone+01};
the spin excitations were found to have a spectral gap $\Delta\approx
1$~meV and a bandwidth of about $1.8$~meV.
It
was shown that the observed spectrum is well described within the
noninteracting magnon approximation, if one assumes the existence of
effective interdimer interactions $J^{\rm eff}_{\vec{\delta}}$ with
$\vec{\delta}=(1,0)$, $(0,1)$, $(1,1)$, $(1,-1)$, $(2,0)$, and
$(0,2)$, with $J_{0}\simeq 2.33$~meV, $J^{\rm eff}_{(1,0)}\simeq 0.44$~meV, $J^{\rm
  eff}_{(0,1)}\simeq 0.23$~meV, $J^{\rm eff}_{(1,1)}\simeq J^{\rm
  eff}_{(1,-1)}\simeq -0.084$~meV, $J^{\rm eff}_{(2,0)}\simeq
-0.073$~meV, and $J^{\rm eff}_{(0,2)}\simeq -0.047$~meV. \cite{comment1}
In this approximation, the magnon dispersion takes the
form
\begin{equation}
\label{e-nonint}
\varepsilon(\vec{k})=\{ J_{0}^{2} +2J_{0}B_{\vec{k}} \}^{1/2},
\quad B_{\vec{k}}=\sum_{\vec{\delta}}
J^{\rm eff}_{\vec{\delta}}\cos(\vec{k}\cdot\vec{\delta})
\end{equation}
and the quasiparticle operators $b_{\vec{k}}$ are connected to the
dimer triplet-creating operators $t_{\vec{k}}$ through the standard
Bogoliubov transformation
$t_{\vec{k}}=u_{\vec{k}}b_{\vec{k}}+v_{\vec{k}}b^{\dag}_{\vec{-k}}$,
where the coefficients $u_{\vec{k}}=\cosh x_{\vec{k}}$,
$v_{\vec{k}}=\sinh x_{\vec{k}}$, with $\tanh
(2x_{\vec{k}})=-B_{\vec{k}}/(J_{0}+B_{\vec{k}})$ and
$\mbox{sign}(v_{\vec{k}})=-\mbox{sign} (B_{\vec{k}})$.

The four-magnon
interaction term $\widehat{H}_{4}$ leads in fact to a renormalization
of the amplitudes $J^{\rm eff}_{\vec{\delta}}$  in $\widehat{H}_{2}$
and can be ignored as far as one uses the values of the effective couplings   $J^{\rm
  eff}_{\vec{\delta}}$ obtained from a comparison with the
experimental data,
since the above renormalization is already included in this way. The
same argument applies to the hardcore constraint which, for low
density of virtual triplet pairs, in the
leading approximation can be shown just to renormalize the self-energy
of a magnon.\cite{Kotov+98}

The three-magnon term $\widehat{H}_{3}$
 does not contribute to the dispersion in the first order of
the perturbation theory, and its main physical effect is to
 produce an instability of
 magnons with energies above the two-particle continuum threshold. In
 terms of transformed quasiparticle operators $b_{\vec{k}}$ this
 interaction can be written as
\begin{eqnarray}
\label{h3c}
 \widehat{H}_{3}&=&\sum_{(xyz)}\sum_{12} M_{12}\; b^{\dag}_{1+2,z}\;
b^{\vphantom{\dag}}_{1,x}\; b^{\vphantom{\dag}}_{2,y}  +\cdots+\mbox{h.c.},\nonumber \\
 M_{12}&=&\sum_{\vec{\delta}} F_{\vec{\delta}} \Big\{
(u_{1+2}-v_{1+2})(v_{1}u_{2}-v_{2}u_{1})\nonumber\\
&\times& \sin(\vec{k}_{1}\cdot\vec{\delta}+\vec{k}_{2}\cdot\vec{\delta})
\\
&-&(u_{1}-v_{1})(u_{1+2}u_{2}+v_{1+2}v_{2})\sin(\vec{k}_{1}\cdot\vec{\delta}) \nonumber\\
&+&(u_{2}-v_{2})(u_{1+2}u_{1}+v_{1+2}v_{1})\sin(\vec{k}_{2}\cdot\vec{\delta})
\Big\}, \nonumber
\end{eqnarray}
where $(xyz)$ denotes a cyclic
summation over the triplet component indices,  $(1,2)\equiv
\vec{k}_{1,2}$, and the remaining terms describing three-particle
creation are omitted for clarity. The decay rate
$\Gamma(\vec{p})$ of a
magnon with the wave vector $\vec{p}$
according to the Fermi golden rule is
\begin{equation}
\label{Gamma}
\Gamma(\vec{p})=2\pi\sum_{\vec{k}}
|M_{\vec{k},\vec{p}-\vec{k}}|^{2}\delta\big(
\varepsilon(\vec{p})-\varepsilon(\vec{k}) -\varepsilon(\vec{p}-\vec{k})\big).
\end{equation}
After
performing the delta-function integration, the latter expression is
reduced to a one-dimensional quadrature which can be computed numerically.

In Ref.\ \onlinecite{MikeskaMueller02} the magnon dispersion in PHCC
was analyzed on the basis of the fourth-order cluster expansion
calculation, and it was concluded that a good fit to the data of Ref.\
\onlinecite{Stone+01} can be achieved even without including the links
with $\vec{\delta}=(2,0)$ or $(0,2)$ (for those latter links, the
existence of exchange paths seems questionable).  Since we are not
taking into account higher-order effects, we will adopt the coupling
values $J^{\rm eff}_{\vec{\delta}}$ proposed in Ref.\
\onlinecite{Stone+01} for the sake of consistency; one may argue
that this is not going to introduce any serious problem since the role
of $J^{\rm eff}_{\vec{\delta}}$ in our calculation here is just to
mimic the correct behavior of $\varepsilon(\vec{k})$ and the
Bogoliubov factors $u_{\vec{k}}$, $v_{\vec{k}}$.

Another conclusion of Ref.\ \onlinecite{MikeskaMueller02} was
that the links with $\vec{\delta}=(1,1)$ and $(1,-1)$ are fully due to
the ``diagonal'' couplings of the $J^{12}$ and $J^{21}$ type,
respectively,
while the other interdimer links have symmetric couplings
in the sense that $J^{12}_{\vec{\delta}}=J_{\vec{\delta}}^{21}$
and $J^{11}_{\vec{\delta}}=J_{\vec{\delta}}^{22}$, so that
all $F_{\vec{\delta}}=0$ except $F_{(1,-1)}\approx -F_{(1,1)}\approx 0.09$~meV.

\begin{figure}[tb]
\includegraphics[width=75mm]{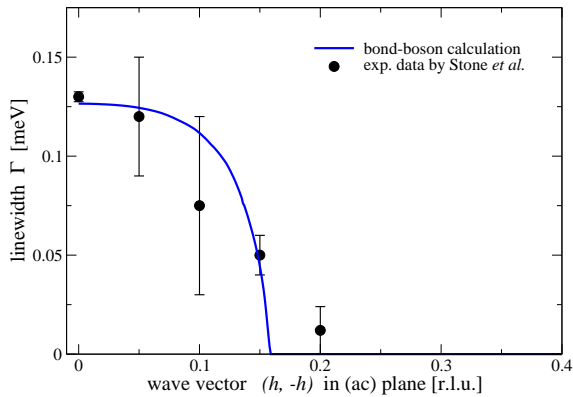}
\caption{\label{fig:gamma} Magnon linewidth $\Gamma(\vec{p})$ in PHCC
as a function of the wave vector $\vec{p}=(h,-h)$ [r.l.u.] in the
$(ac)$ plane. The solid line displays the result of the bond-boson
calculation (see text) and the symbols correspond to experimental data
of Stone et al. \protect\cite{Stone+05}  }
\end{figure}

We have fitted the experimental data \cite{Stone+05} for
$\Gamma$ as a function of the wave vector $\vec{p}$ in the $(ac)$ plane, using the
amplitudes $F_{\vec{\delta}}$ as fitting parameters. If one assumes
that all the individual exchange couplings are antiferromagnetic, a
natural constraint arises
that the absolute value of the amplitude $F_{\vec{\delta}}$ cannot exceed
the absolute value of the corresponding $J^{\rm eff}_{\vec{\delta}}$.
It turns out that the observed magnitude of the linewidth
\cite{Stone+05} (the maximum of $\Gamma \simeq 0.13$~meV is reached at
$\vec{p}=(0,0)$) cannot be achieved if one adopts the coupling pattern
proposed in Ref.\ \onlinecite{MikeskaMueller02}: the contribution of
$F_{(1,1)}$ and $F_{(1,-1)}$ alone is definitely insufficient (with
$\Gamma(0)$ one order of magnitude below the
observed value). The linewidth is  rather insensitive to the
value of $F_{(0,1)}$: even at $F_{(0,1)}=J^{\rm eff}_{(0,1)}$ the
corresponding contribution to $\Gamma(\vec{p}=0)$ is of the order of
$10^{-4}$~meV. Similarly, inclusion of longer links like $F_{(2,0)}$
and $F_{(0,2)}$ does not help, since at their ``full
strength'' (i.e., at $F_{\vec{\delta}}=J^{\rm eff}_{\vec{\delta}}$)
they yield the contributions to $\Gamma(0)$ of the order of several
$\mu$eV. The only remaining way to explain the observed values of
$\Gamma$ is to assume that $F_{(10)}\not=0$; indeed, as shown in Fig.\
\ref{fig:gamma}, a reasonable fit of the experimental data is achieved
with $F_{(1,0)}=0.7 J^{\rm eff}_{(1,0)}$, $F_{(1,1)}=-J^{\rm
eff}_{(1,1)}$, and $F_{(1,-1)}=J^{\rm eff}_{(1,-1)}$. The shown fit
is not unique: the signs of $F_{\vec{\delta}}$ were
chosen to provide the maximum enhancement of the total amplitude of
the matrix element $M_{12}$, and the amplitudes $F_{(2,0)}$ and
$F_{(0,2)}$ for ``questionable'' links were dropped altogether, so
one should view this result as a sort of ``lower bound'' for
$F_{(1,0)}$. Nevertheless, our results are a clear indication that the
interdimer interactions along the $(1,0)$ link in the $(ac)$ plane are
asymmetric with respect to the spin index interchange
$1\leftrightarrow 2$.  It is worth noting that the signs of
$F_{(1,1)}$ and $F_{(1,-1)}$ agree with the pattern proposed in Ref.\
\onlinecite{MikeskaMueller02}.

In PHCC, the crystal symmetry makes all $J^{11}$ and
$J^{22}$ equal, so the only asymmetry giving rise
to nonzero amplitudes $F_{\vec{\delta}}$ can be the difference between
$J^{12}_{\vec{\delta}}$ and $J^{21}_{\vec{\delta}}$.
The existence of a large $F_{\vec{\delta}}$ for the strongest $(1,0)$
link implies that the magnitudes of the individual couplings
$J^{\sigma\sigma'}_{(1,0)}$ are considerably larger than the
effective ``dimer'' exchange constant $J\equiv J^{\rm
  eff}_{(1,0)}\simeq 0.44$~meV. For instance, if one assumes for simplicity that
$J^{12}_{(1,0)}=0$ then $J^{11}_{(1,0)}=J^{22}_{(2,0)}\simeq 1.7
J$ and $J^{21}_{(1,0)}\simeq 1.4J$; another, although less likely
possibility is that the ``diagonal'' couplings have different signs, $J^{12}_{(1,0)}<0$
and $J^{21}_{(1,0)}>0$.

\begin{figure}[tb]
\includegraphics[width=37mm]{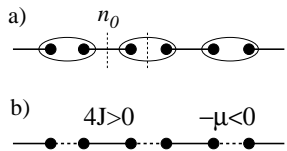}
\caption{\label{fig:hald}
a) representation of spin-$1$ Haldane chain as a spin-$\frac{1}{2}$
chain with
symmetrized  spin-$\frac{1}{2}$ degrees of freedom (denoted by the
ovals); b) at the mean-field level the symmetrization is equivalent to
a ferromagnetic ``Hund's rule'' interaction $-\mu$.
}
\end{figure}

\emph{Magnon decay in $S=1$ Haldane chains.--}
Let us now turn to the $S=1$ spin chain. Indeed, this model can also
be viewed as a chain of $S=\frac{1}{2}$ dimers with antiferromagnetic
intradimer exchange $4J$ and an additional local constraint
symmetrizing the edge spins of every two neighboring dimers, and thus
coupling them into an effective spin-$1$, see Fig.\
\ref{fig:hald}. This constraint can be taken into account `on average'
at a mean-field level by adding the Lagrange multiplier term of the form
$-\mu \sum_{n} (\vec{S}_{1,n+1}\cdot \vec{S}_{2,n} -\frac{1}{4})$,
playing the role of a ferromagnetic ``Hund's rule'' interdimer
interaction which has to be determined self-consistently. It is then
obvious that the interaction between the dimers is asymmetric and the
effective Hamiltonian will contain the term of the form (\ref{h3})
with a single $F_{1}=\frac{1}{2}\mu$. It is easy to see that this term
is consistent with all symmetries of the initial Hamiltonian, e.g., it
satisfies the
reflection symmetry about the $n_{0}$-th bond, which in the bond-boson
language  corresponds to the
transformation $t_{n,\alpha}\mapsto -t_{2n_{0}-n,\alpha}$, as well as
the symmetry of reflection about the $n_{0}$-th ``site'' (i.e., the interdimer
bond to the right of the $n_{0}$-th dimer, see Fig.\
\ref{fig:hald}a), which corresponds to $t_{n,\alpha}\mapsto -t_{2n_{0}+1-n,\alpha}$.

%
%

Then, a simple golden rule calculation yields the magnon linewidth in the Haldane
chain as a function of the
its momentum $p$:
\begin{equation}
\label{lw-hald1}
\Gamma(p)=|M_{k_{0},p-k_{0}}|^{2}/
|\varepsilon'_{k_{0}}-\varepsilon'_{p-k_{0}}|,
\end{equation}
where $k_{0}$ is the solution of the equation
$\varepsilon_{k_{0}}+\varepsilon_{p-k_{0}}=\varepsilon_{p}$, and
$\varepsilon_{k}=2(4J^{2}+\mu J\cos k )^{1/2}$ is the single-magnon
dispersion.  Using the known value of the Haldane gap $\Delta\simeq
0.4J$, one can fix the renormalized value of $\mu$ to be slightly
below $4J$. According to (\ref{h3}), the matrix element has the
structure $M_{k,p-k}=f_{p}(k)-f_{p}(p-k)$, so the linewidth becomes
zero at the wave vector threshold $p=p_{c}$, defined by
$\varepsilon(p_{c})=2\varepsilon(p_{c}/2)$ (here $p_{c}\in
[\pi,2\pi]$); it is exactly the point of crossing with the the
two-particle continuum. Above the continuum threshold, in the vicinity
of $p_{c}$ the linewidth grows as a square root:
\begin{equation}
\label{lw-hald2}
\Gamma(p)\simeq \frac{2|f'(p_{c}/2)|^{2}}{|\varepsilon''(p_{c}/2)|}
\Big|
\frac{\varepsilon'(p_{c})-\varepsilon'(p_{c}/2)}{\varepsilon''(p_{c}/2)}\Big|^{1/2}
\sqrt{p-p_{c}}.
\end{equation}
It is notable that two-magnon decay
processes, strictly forbidden in the Haldane chain within the
effective continuum field theory, nevertheless survive in the full
lattice description.

Using the above simple estimate for $\mu$, one can see that the
magnon linewidth in the Haldane chain grows fast to rather large
values, e.g., at $p=0$ one obtains  $\Gamma\simeq 3.4J$, to be
compared to $\varepsilon(p=0)\simeq 5.6 J$. Such a large value of
the linewidth arises because the effective ``interdimer''
interaction $\mu$ is not small compared to the ``intradimer''
coupling $4J$.
An experimental observation in that case would show
a very fast disappearance of quasiparticle excitation above the
continuum threshold, similar to the picture observed
recently \cite{Masuda+05} in a spin ladder compound $\rm
(CH_{3})_{2}CHNH_{3}CuCl_{3}$. However, one should bear in mind that
the behavior of linewidth in the vicinity of the threshold
depends on the details of the wave vector dependence of the matrix
element, which cannot be accounted for in the simple qualitative
description sketched above: for instance, if $f'(p_{c}/2)$ is anomalously small  for some
reason, then one has to take into account higher derivatives of $f(p)$ and
(\ref{lw-hald2}) gets replaced
by a sum of competing terms, $\Gamma(p)\propto \sum_{n=1} A_{n}(p-p_{c})^{n}$.
Recently, a finite magnon linewidth slowly emerging at $p_{c}\simeq 0.6\pi$
was observed \cite{Zaliznyak+01} in the $S=1$ chain material $\rm
CsNiCl_{3}$.
It would be
interesting if future experiments could provide more information on
two-magnon decay in Haldane spin chain.

\emph{Summary.--} To conclude, we have shown that elementary
excitations in a gapped dimerized spin system can become unstable
above the two-particle continuum threshold, and that the dependence of
the magnon linewidth on the wave vector can be used as an additional
input (complementing the magnon dispersion) to extract the information
on the exchange interactions pattern. On the basis of fitting the
recent results \cite{Stone+05} on the magnon linewidth in the
quasi-two-dimensional spin dimer material PHCC, we conclude that
interdimer couplings along the $a$ axis in this material are strongly
asymmetric, contrary to the previously assumed \cite{MikeskaMueller02}
pattern. We argue that three-magnon interaction, which makes possible
the instability of a single-magnon excitation against the decay into
two particles, is generically present in other gapped spin systems
such as spin-$1$ Haldane chains.

\emph{Acknowledgments.--} We thank C. Broholm, T. Senthil, and
I. Zaliznyak for useful discussions.
This research was supported by NSF Grant
DMR-0537077. AK is supported by the Heisenberg Program Grant No.\
KO~2335/1-1 from Deutsche Forschungsgemeinschaft.


\begin{thebibliography}{10}
\bibitem{comment1} The constants $B_{0}$, $B_{\vec{\delta}}$ given in
Table II of Ref.\ [\protect\onlinecite{Stone+01}] are connected to our
notation through $B_{0}=J_{0}^{2}$, $B_{\vec{\delta}}=2 J_{0} J^{\rm
eff}_{\vec{\delta}}$.
\end{thebibliography}
\end{document}